\documentclass[a4paper]{jpconf}
\usepackage{graphicx}
\usepackage{color}
\usepackage{ulem}
\usepackage{iopams}

\newcommand{\DI}{\ensuremath{D_\mathrm{I}}}

\newcommand{\Tc}{\ensuremath{T_\mathrm{c}}}
\newcommand{\Ns}{\ensuremath{N_\mathrm{s}}}

\newcommand{\valpha}{\ensuremath{\vec{\alpha}}}
\newcommand{\vQ}{\ensuremath{\vec{Q}}}

\begin{document}
\title{Loop algorithm for classical antiferromagnetic Heisenberg models with biquadratic interactions}

\author{Hiroshi Shinaoka$^{1,2}$, Yusuke Tomita$^1$ and Yukitoshi Motome$^3$}
\address{$^1$Institute for Solid State Physics, University of Tokyo, Kashiwanoha, Kashiwa, Chiba, 277-8581, Japan}
\address{$^2$Present address: Nanosystem Research Institute, AIST, Tsukuba 305-8568, Japan}
\address{$^3$Department of Applied Physics, University of Tokyo, 7-3-1 Hongo, Bunkyo-ku, Tokyo 113-8656, Japan}

\ead{h.shinaoka@aist.go.jp}

\begin{abstract}
Monte Carlo simulation using the standard single-spin flip algorithm often fails to sample over the entire configuration space at low temperatures for frustrated spin systems.
A typical example is a class of spin-ice type Ising models.
In this case, the difficulty can be avoided by introducing a global-flip algorithm, the loop algorithm.
Similar difficulty is encountered in $O$(3) Heisenberg models in the presence of biquadratic interaction.
The loop algorithm, however, is not straightforwardly applied to this case, since the system does not have a priori spin-anisotropy axis for constructing the loops.
We propose an extension of the loop algorithm to the bilinear-biquadratic models. 
The efficiency is tested for three different ways to flip spins on a loop in Monte Carlo simulation.
We show that the most efficient method depends on the strength of the biquadratic interaction.
\end{abstract}

\section{Introduction}
Monte Carlo (MC) simulation is a powerful tool for investigating thermodynamic properties of classical spin models~\cite{Landau-Binder00}.
The standard single-spin flip algorithm is widely used because it is simple and applicable to systems with any type of interactions.
However, it is well known that it suffers from slow relaxation in many cases.
For example, the single-spin-flip dynamics exhibits critical slowing down near a continuous transition temperature where the correlation length diverges.
Such difficulty is avoided by using a nonlocal, global update, such as the Swendsen-Wang cluster algorithm for Ising systems~\cite{Swendsen87} or its Wolff's extension to Heisenberg systems~\cite{Wolff89}.
These cluster algorithms accelerate the relaxation by flipping many spins at once, i.e., updating the spin configuration drastically.

\begin{figure}[!]
 \centering
 \includegraphics[width=.65\textwidth,clip]{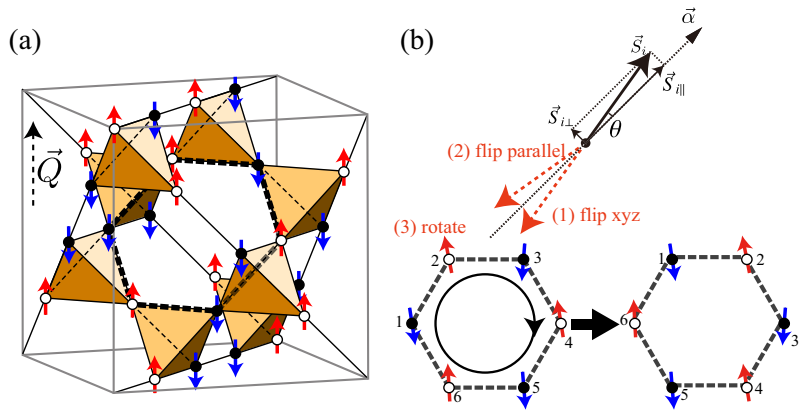}
\caption{\label{Fig1} (a) The pyrochlore lattice composed of a three-dimensional network of corner-sharing tetrahedra. A 16-site cubic unit cell is shown. Spins are denoted by solid arrows. The broken arrow $\vQ$ shows a common axis selected by $b$ in the nematic phase. Black (filled) circles represent spins in the direction of $\vQ$, while white (open) circles represent spins in the opposite direction. 
The spin configuration is an example of the spin-ice type states in which the `two-up two-down' local constraint is satisfied in every tetrahedron.
The hexagon with a bold dashed line denotes one of the shortest loops on which a flip of all spins (black and white) transforms the ice-rule state to another ice-rule state. (b) Three different ways to flip black and white: (1) \textit{flip xyz}, (2) \textit{flip parallel}, and (3) \textit{rotate}. See the text for details.}
 \label{fig:1}
\end{figure}

Besides the critical slowing down, the single-spin flip algorithm often suffers from dynamical freezing at low temperature ($T$) when the ground state has macroscopic degeneracy under the influence of geometrical frustration.
A typical example is an antiferromagnetic Ising model on a pyrochlore lattice.
The pyrochlore lattice is a three-dimensional frustrated structure given by a corner-sharing network of tetrahedra, as shown in Fig.~\ref{fig:1}(a).
When the antiferromagnetic exchange interaction is limited to nearest-neighbor sites, no long-range ordering occurs down to zero $T$ and the ground state has macroscopic degeneracy~\cite{Anderson56}.
The degenerate manifold is identified by a collection of local constraints enforcing two spins pointing up and two spins pointing down in every tetrahedron, as exemplified in Fig.~\ref{fig:1}(a).
This `two-up two-down' constraint is called the ice rule because of an analogy to the constraint on positions of protons in hexagonal ice~\cite{Bernal33,Pauling35}.
Similar situation was recently discovered in the so-called spin-ice systems, 
in which the Ising-like spins point along the local $\langle 111 \rangle$ axes and 
interact with each other by ferromagnetic exchange interaction and dipole interaction~\cite{Harris97,Ramirez99}.
Because the degenerate `ice-rule' configurations are separated by large energy barriers of the order of the dominant interaction scale $J$, the single-spin flip does not work at low $T \ll J$. 
The difficulty is avoided by introducing a global flip called the loop flip, in which one reverses all Ising spins on a specific closed loop passing through tetrahedra~\cite{Rahman72, Yanagawa79, Barkema98,Melko01, Melko04}; the loop is chosen so that the spins are up and down (inward and outward in the spin-ice problem) alternatively along the loop [see the hexagon in Fig.~\ref{fig:1}(a) as an example].
This loop flip enables to transform an ice-rule state to another ice-rule state bypassing the energy barriers.

The difficulty remains even when the Ising discreteness is relaxed and spins can fluctuate, as long as the ground-state manifold retains a multivalley structure with large energy barriers. 
Such situation is seen in variants of pyrochlore antiferromagnets, such as classical Heisenberg models in the presence of the single-ion easy-axis anisotropy~\cite{Champion02,Shinaoka10a} and the biquadratic interaction~\cite{Shannon10}.
In contrast to the Ising case, however, it is nontrivial how to define the loop with alternating spins.
Moreover, the loop flip procedure is not unique in the Heisenberg spin case because of the continuous degrees of freedom.
Recently, the authors extended the loop algorithm to the Heisenberg spin systems with single-ion anisotropy by defining `colors' of spins as black and white, which is a natural generalization of up and down in the antiferromagnetic Ising case, in terms of the projection of spins on the anisotropy axis~\cite{Shinaoka10a}.
We tested different ways to flip spins on a formed loop and showed that the efficiency strongly depends on the method. 
In the models with the biquadratic interaction, however, because of the $O(3)$ spin rotational symmetry, we have no explicit anisotropy axis to project spins on.
Therefore, an extension of the loop algorithm to this class of models is not straightforward.

In this paper, we develop an extension of the loop algorithm to classical antiferromagnetic Heisenberg spin systems with biquadratic interaction in which the spin-ice type manifold emerges at low $T$. 
We propose a way to define the projection axis for constructing loops and test three different ways to flip a formed loop.
We apply the algorithm to a nearest-neighbor bilinear-biquadratic model, and compare the efficiency of the three methods.
Interested readers are referred to Ref.~\cite{Shinaoka10b} for further application of the present algorithm to a bond-disordered bilinear-biquadratic model.

\section{Extension of the loop algorithm to bilinear-biquadratic spin systems}
\subsection{Model}
In this section, we extend the loop algorithm to classical antiferromagnetic Heisenberg models with biquadratic interactions.
We start with a Hamiltonian of a simple form:
\begin{equation}
\mathcal{H}=\sum_{\langle i,j\rangle} \left\{J\left(\vec{S}_i \cdot \vec{S}_j\right) - b\left(\vec{S}_i \cdot \vec{S}_{j}\right)^2 \right\}, \label{eq:ham-h}
\end{equation}
where $\vec{S}_i$ denotes a classical Heisenberg spin at site $i$ on the pyrochlore lattice [Fig.~\ref{fig:1}(a)] (we take $|\vec{S}_i|=1$) and $b$ is the coupling constant of the biquadratic interaction. 
Here we consider $b>0$ which favors collinear spin configurations. We note that such `ferro'-type biquadratic interaction originates in the spin-lattice coupling as well as quantum and thermal fluctuations. We consider the antiferromagnetic exchange interaction $J>0$, and take the energy unit as $J=1$.
The sum runs over the nearest-neighbor bonds of the pyrochlore lattice.
The following algorithm is applicable to more general spin-ice type models on other frustrated lattices with farther-neighbor or bond-dependent interactions, and site-dependent anisotropy.

When $b=0$, the model given by Eq.~(\ref{eq:ham-h}) exhibits no long-range ordering down to $T=0$, and the ground state is given by a collection of local constraints that enforces the sum of $\vec{S}_i$ to be zero in every tetrahedron. 
Consequently, the ground-state manifold has continuous macroscopic degeneracy~\cite{Reimers92,Moessner98a,Moessner98b}.
For $b>0$, the present model exhibits a weak first-order transition at $\Tc \sim b$ to a nematic state in which spins spontaneously select a common axis $\vQ$ without selecting their directions on it~\cite{Shannon10}.
Hence, the ground state for $b>0$ is identified by a collection of spin-ice type local constraints: in every tetrahedron, two out of four spins are aligned parallel to each other and the other two are antiparallel to them --- `two-up two-down' configuration [see Fig.~\ref{fig:1}(a)].
The ground-state manifold develops a multivalley structure whose minima correspond to different spin-ice-type configurations separated by large energy barriers of the order of $b$ and $J$.  

\subsection{Algorithm}
In the previous paper, the authors developed an extension of the loop algorithm to be applicable to a family of classical Heisenberg antiferromagnets with single-ion anisotropy, in which the single-spin flip algorithm suffers from slow relaxation due to the formation of the spin-ice type manifold~\cite{Shinaoka10a}. 
In the extended loop algorithm, 
(i) we first project all the spins onto the anisotropy axis to assign black and white colors, 
(ii) next, construct a loop of alternating black and white, and 
(iii) flip all the spins on the loop. 
For the present bilinear-biquadratic model, 
a similar difficulty from slow relaxation is anticipated 
because the low-$T$ state develops the spin-ice type degeneracy as mentioned above. 
However, it is not straightforward to apply the extended loop algorithm 
since the present model retains $O$(3) spin rotational symmetry and 
does not have any explicit anisotropy axis to project the spins on. 
It is necessary to deduce the common axis $\vQ$ selected by $b$ for each MC sample.

Here, we propose the following procedure to define the projection axis.
We first pick up a set of $N_\mathrm{T}$ tetrahedra \{$\mathcal{T}_m$\} ($m=1,\cdots,N_\mathrm{T}$) randomly from the whole system.
Starting from an initial guess $\valpha_0$ [we take $\valpha_0 = (0,0,1)$], the normalized projection axis $\valpha$ is obtained iteratively by
\begin{eqnarray}
\valpha_{n+1} \propto \sum_{i \in \{ \mathcal{T}_m \}} \mathrm{sign} (\vec{S}_{i} \cdot \valpha_n) \vec{S}_{i}.
\end{eqnarray}
Here the sum is taken over all spins belonging to the selected tetrahedra \{$\mathcal{T}_m$\}, and $n~(=0,1,\cdots,n_\mathrm{max}-1)$ is the index of the iteration.
For larger $N_\mathrm{T}$ and $n_\mathrm{max}$, the resultant $\valpha=\valpha_{n_\mathrm{max}}$ gives a better approximation of $\vQ$. In practice, we take $N_\mathrm{T} = 24$ and $n_\mathrm{max}=6$ in the following calculations for the system size with $\Ns=16\times8^3$ spins. We confirm that the loop flip is efficiently performed for these conditions, 
as demonstrated below.
Once $\valpha$ is defined in this way, the step (ii) and (iii) are done in the same way as in Ref.~\cite{Shinaoka10a}.
It should be noted that, to ensure the detailed balance, loops must be constructed avoiding the tetrahedra included in \{$\mathcal{T}_m$\} as well as defect tetrahedra in which the ice rule is violated: Otherwise, the loop flip becomes irreversible because the flip changes $\valpha$.

In the extended loop algorithm, the loop flip procedure to reverse all colors on a loop is not unique because spins can thermally fluctuate~\cite{Shinaoka10a}.
To choose an efficient method, careful consideration on the energy change is necessary.
In Ref.~\cite{Shinaoka10a}, the authors tested two different ways: (1) \textit{flip xyz} and (2) \textit{flip parallel}.
In \textit{flip xyz}, all three Cartesian components of $\vec{S}_i$ are reversed as $\vec{S}_i \rightarrow - \vec{S}_i$, while in \textit{flip parallel}, only parallel components $\vec{S}_{i\parallel}$ are reversed as $\vec{S}_i \rightarrow \vec{S}_i  - 2 (\vec{S}_i \cdot \valpha) \valpha$ [see Fig.~\ref{fig:1}(b)].
The previous study revealed that, in the case of the single-ion anisotropy, only the acceptance rate for \textit{flip parallel} can become one (rejection free) in the limit of $T\rightarrow 0$; 
the acceptance rate for \textit{flip xyz} converges to a finite value less than unity as $T\to 0$ because of the effect of thermal fluctuations on the transverse component of spins~\cite{Shinaoka10a}. 
In addition to these two updates, in this paper, we introduce another way to reverse colors, i.e., (3) \textit{rotate}. In \textit{rotate}, one translates every spin to the neighboring site on the loop simultaneously in the direction in which the loop was formed [see Fig.~\ref{fig:1}(b)].
In the next section, we try the three different methods and demonstrate that the most efficient method depends on the model parameter $b$.

\section{Benchmark}
In this section, we apply the extended loop algorithm to the model given by Eq.~(\ref{eq:ham-h}). We demonstrate the efficiency of loop flips in MC simulations and compare the efficiency of the three methods.
In the following, we show the MC results for the systems size with $\Ns=16\times8^3$ spins under periodic boundary conditions.
To retain the ergodicity, we use the loop flip together with the single-spin flip. One MC step consists of single-spin flips, followed by loop flips with either \textit{flip xyz}, \textit{flip parallel}, or \textit{rotate}.
In the single-spin flips, we randomly choose a new spin state on the unit sphere for each spin following a procedure proposed by Marsaglia~\cite{Marsaglia72}.
\begin{figure}[!]
 \centering
 \includegraphics[width=.7\textwidth,clip]{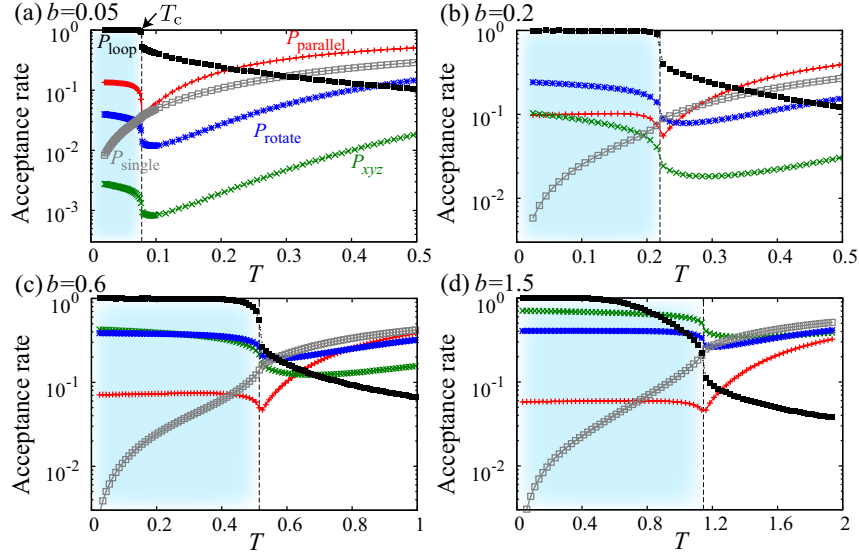}
 \caption{Temperature dependences of the acceptance rates of the single-spin flip ($P_{\mathrm{single}}$), the probability of formation of closed loops ($P_{\mathrm{loop}}$), the acceptance rates of flip of a formed loop by \textit{flip xyz} ($P_{xyz}$), \textit{flip parallel} ($P_{\mathrm{parallel}}$), and \textit{rotate} ($P_\mathrm{rotate}$). The data are calculated at (a) $b=0.05$, (b) $0.2$, (c) $0.6$, and (d) $1.5$. The vertical broken lines denote the nematic transition temperature $\Tc$ estimated by the peak position in the specific heat (not shown).}
 \label{fig:2}
\end{figure}
\begin{figure}[!]
 \center
 \includegraphics[width=.425\textwidth,clip]{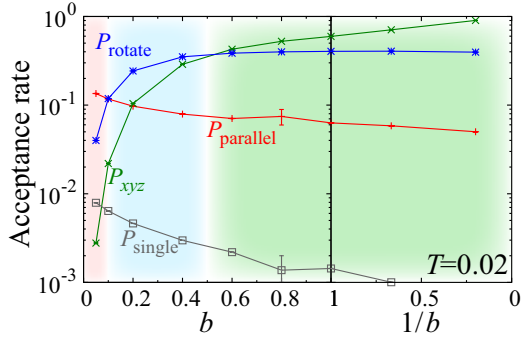}\hspace{2pc}
 \begin{minipage}[b]{14pc}
	 \caption{$b$ dependence of $P_{\mathrm{parallel}}$, $P_{xyz}$, $P_{\mathrm{rotate}}$, and $P_{\mathrm{single}}$ at $T=0.02$.}
 \end{minipage}
 \label{fig:3}
\end{figure}

At low $T$ compared to $b$ and $J$, spin configurations are enforced to satisfy the `two-up two-down' ice rule, and the acceptance rate of the single-spin flip, $P_{\mathrm{single}}$, is suppressed. This is demonstrated in Fig.~\ref{fig:2}. On the other hand, the probability that a closed loop is successfully formed, $P_\mathrm{loop}$, steeply increases below the nematic transition temperature $\Tc \sim b$, indicating that almost all tetrahedra start to follow the ice rule below $\Tc$. 
[For $b>J$, the ice rule is weakly violated in the range of $T \gtrsim J$ even below $\Tc$, but gradually satisfied for $T \lesssim J$, as exemplified in Fig.~\ref{fig:2}(d).]
The acceptance rate of flips of a formed loop also increases below $\Tc$ and remains finite as $T \to 0$; here, $P_{xyz}$, $P_{\mathrm{parallel}}$, and $P_\mathrm{rotate}$ are the rate for \textit{flip xyz}, \textit{flip parallel}, and \textit{rotate}, respectively.
The total acceptance rate of the loop flip is given by the product as $P_\mathrm{loop} \times P_{xyz}$, $P_\mathrm{loop} \times P_{\mathrm{parallel}}$, and $P_\mathrm{rotate} \times P_{\mathrm{parallel}}$ for each method.
Hence the acceptance rate of the loop flip sharply increases at $T < \Tc$, compensating the decrease of $P_{\mathrm{single}}$. 
These $T$ dependences are qualitatively the same for the wide range of $b$, as shown in Fig.~\ref{fig:2}.

As demonstrated in the previous study for the models with single-ion anisotropy~\cite{Shinaoka10a}, the efficiency of the loop flip at low $T$ depends on the method.
Furthermore, in the present case with the biquadratic interaction $b$, the efficiency at low $T$ strongly depends on $b$. The result is presented in Fig.~\ref{fig:3}.
For small $b < 0.1$, \textit{flip parallel} is most efficient, while it is taken over by \textit{rotate} in the intermediate range of $0.1 < b < 0.5$, and finally by \textit{flip xyz} for $b > 0.5$. 

The difference of the efficiency is understood by the following consideration on the energy change by the flips.
Considering a given state at a finite $T$ well below $\Tc$, its energy measured from the ground-state energy is given by $E=E_J + E_b$, where $E_J$ and $E_b$ are the energies corresponding to the first and second terms in Eq.~(\ref{eq:ham-h}), respectively. 
Both $E_J$ and $E_b$ are of the order of $T$ at low $T$.
The three loop flips change the two contributions in different ways. 
The \textit{flip xyz} changes $E_J$ by a certain fraction $\Delta E_J$ but conserves $E_b$.
Meanwhile, \textit{flip parallel} and \textit{rotate} change both of $E_J$ and $E_b$ by $\Delta E_J$ and $\Delta E_b$, respectively. 
First we consider the large $b$ limit, where $E \simeq E_b\propto T$ and $E_J/E_b \propto b^{-1}$.
For \textit{flip xyz} in which $\Delta E_b = 0$, we obtain $\Delta E/T = \Delta E_J/T \propto b^{-1} \to 0$ as $b \to \infty$. 
Since the acceptance rate is given by $\exp\left(-\Delta E/T\right)$, this consideration gives $\lim_{b\to +\infty} P_{xyz} \to 1$, that is, \textit{flip xyz} becomes rejection free as $b \to +\infty$.
This is consistent with the behavior in Fig.~\ref{fig:3}.
On the contrary, \textit{flip parallel} and \textit{rotate} cannot become rejection free for $b \to \infty$ because they change $E_b$ by $O(T)$; $P_\mathrm{parallel}$ and $P_\mathrm{rotate}$ converge to finite values less than unity at $b\to \infty$, respectively.
Note that \textit{rotate} does not change a half of the nearest-neighbor bond energies.
This may account for why $P_\mathrm{rotate} > P_\mathrm{parallel}$ at $b\to +\infty$.

For smaller $b$, spins deviate from the common axis $\vQ$ with larger angle. 
Considering that $\vQ$ is set by $E_b \sim T$ and $E_b$ is proportional to $b\theta^2$ ($\theta$ is a typical deviation angle), we obtain $\theta = O(\sqrt{T/b})$ for $b\ll J$~\cite{comment}.
Since \textit{flip parallel} changes $E_J$ and $E_b$ by $O(J\theta^4)$ and $O(b\theta^2)$, respectively~\cite{Shinaoka10a}, we obtain $\Delta E/T = \Delta E_b/T = O(1)$ at low $T\ll \Tc$ for \textit{flip parallel}.
This indicates that $P_\mathrm{parallel}$ does not vanish even at $b\to 0$ in the nematic phase.
Meanwhile, the other two methods change the energy as $\Delta E/T =\Delta E_J/T=O(J\theta^2/T)=O(J/b)$ at low $T$ and $b\ll J$~\cite{Shinaoka10a}.
This suggests that their acceptance rates vanish as $b\to 0$ in contrast to \textit{flip parallel}.
Therefore, \textit{flip parallel} becomes most efficient at $b\to 0$.
These considerations are consistent with the numerical results shown in Fig.~3.
In the intermediate regime, i.e, $0.1<b<0.5$, \textit{rotate} is superior to the other two, presumably because of a remnant of the advantage of \textit{rotate} over \textit{flip parallel} at $b \to \infty$.

\section{Summary}
In this paper, we have extended the loop algorithm to the classical antiferromagnetic Heisenberg spin models with biquadratic interaction which have spin-ice type ground-state degeneracy.
The efficiency of the extended loop algorithm has been demonstrated in Monte Carlo simulations. 
We have examined three different ways of loop flips, \textit{flip xyz}, \textit{flip parallel}, and \textit{rotate}, and compared their efficiency.
We have shown that the most efficient method depends on the strength of the biquadratic interaction $b$.
This $b$ dependence has been explained by considering effects of thermal fluctuations on the energy changes by the loop flips.

This work was supported by Grant-in-Aids (No. 19052008), Global COE Program ``the Physical Sciences Frontier'', and the Next Generation Super Computing Project, Nanoscience Program, MEXT, Japan.

\section*{References}

\begin{thebibliography}{10}
\expandafter\ifx\csname url\endcsname\relax
  \def\url#1{{\tt #1}}\fi
\expandafter\ifx\csname urlprefix\endcsname\relax\def\urlprefix{URL }\fi
\providecommand{\eprint}[2][]{\url{#2}}

\bibitem{Landau-Binder00}
Landau D~P and Binder K 2000 {\em A guide to Monte Carlo simulation in
  statistical physics\/} (Cambridge: Cambridge Univ. Press)

\bibitem{Swendsen87}
Swendsen R~H and Wang J~S 1987 {\em Phys. Rev. Lett.\/} {\bf 58} 86--88

\bibitem{Wolff89}
Wolff U 1989 {\em Phys. Rev. Lett.\/} {\bf 62} 361--364

\bibitem{Anderson56}
Anderson P~W 1956 {\em Phys. Rev.\/} {\bf 102} 1008--1013

\bibitem{Bernal33}
Bernal J~D and Fowlers R~H 1933 {\em J. Chem. Phys.\/} {\bf 1} 515--548

\bibitem{Pauling35}
Pauling L 1935 {\em J. Am. Chem. Soc.\/} {\bf 57} 2680

\bibitem{Harris97}
Harris M~J, Bramwell S~T, McMorrow D~F, Zeiske T and Godfrey K~W 1997 {\em
  Phys. Rev. Lett.\/} {\bf 79} 2554--2557

\bibitem{Ramirez99}
Ramirez A~P, Hayashi A, Cava R~J, Siddharthan R and Shastry B~S 1999 {\em
  Nature\/} {\bf 399} 333--335

\bibitem{Rahman72}
Rahman A and Stillinger F~H 1972 {\em J. Chem. Phys.\/} {\bf 57} 4009

\bibitem{Yanagawa79}
Yanagawa A and Nagle J~F 1979 {\em Chem. Phys.\/} {\bf 43} 329

\bibitem{Barkema98}
Barkema G~T and Newman M~E~J 1998 {\em Phys. Rev. E\/} {\bf 57} 1155--1166

\bibitem{Melko01}
Melko R~G, den Hertog B~C and Gingras M~J~P 2001 {\em Phys. Rev. Lett.\/} {\bf 87} 067203

\bibitem{Melko04} Melko R~G and Gingras M~J~P 2004 {\em J. Phys.: Condens. Matter\/} {\bf 16} R1277

\bibitem{Champion02}
Champion J~D~M, Bramwell S~T, Holdsworth P~C~W and Harris M~J 2002 {\em EPL
  (Europhysics Letters)\/} {\bf 57} 93

\bibitem{Shinaoka10a}
Shinaoka H and Motome Y 2010 {\em Phys. Rev. B\/} {\bf 82} 134420

\bibitem{Shannon10}
Shannon N, Penc K and Motome Y 2010 {\em Phys. Rev. B\/} {\bf 81}
  184409--1--184409--24

\bibitem{Shinaoka10b}
Shinaoka H, Tomita Y and Motome Y 2010 (\textit{Preprint}
  \eprint{cond-mat/1010.5625})

\bibitem{Reimers92}
Reimers J~N 1992 {\em Phys. Rev. B\/} {\bf 45} 7287--7294

\bibitem{Moessner98a}
Moessner R and Chalker J~T 1998 {\em Phys. Rev. Lett\/} {\bf 80} 2929--2932

\bibitem{Moessner98b}
Moessner R and Chalker J~T 1998 {\em Phys. Rev. B\/} {\bf 58} 12049--12062

\bibitem{Marsaglia72}
Marsaglia G 1972 {\em The Annals of Mathematical Statistics\/} {\bf 43}
  645--646

\bibitem{comment} In the case of the single-ion anisotropy considered in Ref.~\cite{Shinaoka10a}, we assumed that $\theta\propto T$. It should be corrected as $\theta = O(\sqrt{T/\DI})$ ($\DI$ is the anisotropy). This, however, does not alter the arguments in the previous study.

\end{thebibliography}
\providecommand{\newblock}{}

\end{document}